\documentstyle[12pt]{article} 
\title{Stationary generalized Kerr--Schild spacetimes} 
 
\author{Carlos F. Sopuerta\thanks{Also at Laboratori de F\'{\i}sica 
Matem\`{a}tica.  Societat Catalana de F\'{\i}sica. IEC. Barcelona. 
\newline 
\mbox{}\indent \hspace{2mm} E--mail address: carlos@hermes.ffn.ub.es}  
\\ Departament de F\'{\i}sica Fonamental, 
Universitat de Barcelona\\ Diagonal 647, 08028 Barcelona, Spain.} 
 
\begin{document} 
 
 
\newcommand{\aaa}{{\cal A}} 
\newcommand{\kk}{\mbox{\cal{k}}} 
\newcommand{\mm}{\mbox{\cal{m}}} 
\newcommand{\hh}{{\cal H}} 
\newcommand{\uu}{{\cal U}} 
\newcommand{\vv}{{\cal V}} 
\newcommand{\ww}{{\cal W}} 
\newcommand{\acc}{\mbox{\cal{a}}}         
\newcommand{\qu}{\mbox{\cal{q}}}          
 
\renewcommand{\thesection}{\Roman{section}}

 
\maketitle 
 
\begin{abstract} 
In this paper we have applied the generalized Kerr--Schild  
transformation finding a new family of stationary perfect--fluid  
solutions of the Einstein field equations.  The procedure used combines  
some well--known techniques of null and timelike vector fields, from  
which some properties of the solutions are studied in a  
coordinate--free way. These spacetimes are algebraically special being  
their Petrov types II and D. This family includes all the classical  
vacuum Kerr--Schild spacetimes, excepting the plane--fronted  
gravitational waves, and some other interesting solutions as, for  
instance, the Kerr metric in the background of the Einstein Universe. 
However, the family is much more general and depends on an arbitrary 
function of one variable.   
\end{abstract} 
 
\vspace{4cm} 
 
 
\newpage 
 
 
\section{Introduction} 
\mbox{}\indent 
Since the discovery of the Kerr metric \cite{KERR} until now, the 
Kerr--Schild Ansatz \cite{KESC}-\cite{DEKS} has been a powerful tool  
to find new explicit solutions of the Einstein equations. Roughly  
speaking, this technique consists in generating new solutions from  
the Minkowski spacetime and its null geodesic vector fields, and it  
can be applied to several types of energy--momentum content: vacuum,  
electromagnetic field, pure radiation, etc. With regard to the  
results obtained, we should note that some of the solutions found  
have been of crucial importance in general relativity and they have  
received too much attention, specially in relation to the  
description of black holes or radiating bodies. Well known examples  
are the Kerr--Newman family of spacetimes, the Vaidya metric, etc.   
A complete review of the classical Kerr--Schild transformation can  
be found in reference \cite{KSHM} (chapter 28).  
 
Some generalizations of this generation technique have appeared, so that 
they overcome some of the limitations that the classical Kerr--Schild 
transformation had.  In particular, the energy--momentum tensor of the  
classical Kerr-Schild spacetimes generated have the null vector field  
used in the transformation as a null eigenvector and hence, they cannot  
be perfect--fluid solutions.  A generalization which allows to find  
perfect--fluid spacetimes is the generalized Kerr--Schild (GKS hereafter)  
transformation, which has been studied widely in  
\cite{MASE}--\cite{NAHM}. The GKS transformation is a generation  
technique in which the metrics of the initial and the final spacetimes  
are related by the following expression  
\begin{equation} 
\tilde{g}_{ab} \hspace{2mm} = \hspace{2mm} g_{ab} \hspace{2mm} + 
\hspace{2mm} 2\,\hh\,\ell_{a} \ell_{b} , \label{gkst} 
\end{equation} 
where the starting metric $g_{ab}$, which we shall call {\it seed  
metric}, is any metric, $\vec{\ell}$ is a null vector field for the  
metric $g_{ab}$, and $\hh$ is a scalar field. In the case in which  
$\vec{\ell}$ is geodesic, an illuminating result was proved in  
\cite{MASE,SENO} (see also \cite{MPSE,SESO,SOPU}). It states that the  
energy--momentum tensor of the GKS spacetimes has $\vec{\ell}$ as an  
eigenvector if and only if $\vec{\ell}$ is an eigenvector for the  
energy--momentum tensor of the seed spacetimes.  Therefore, if we are  
interested in finding perfect--fluid spacetimes, we should start with a  
seed spacetime whose energy--momentum tensor does not have   
null eigenvectors. In this line, a particularization of the GKS  
transformation for the search of perfect--fluid solutions was developed  
in \cite{MASE,SENO}, where the seed spacetime was taking to be  
conformally--flat.  From the application of this technique to several  
cases, new perfect--fluid solutions of Einstein equations have been  
obtained. In some of these applications the solutions represent  
inhomogeneous cosmological models while in others, they are static and  
stationary spacetimes (see \cite{MASE}--\cite{SESO}, \cite{TAUB} and  
\cite{VAID}).

The subject of this paper is to apply the GKS transformation, in  
combination with techniques of null and timelike vector fields, in order  
to find new perfect--fluid solutions of Einstein's equations. The  
result found is a wide new family of stationary perfect--fluid  
spacetimes.

In the development of this work we follow the spirit of the classical  
Kerr--Schild transformation, where the seed spacetime was very simple  
(the Minkowski spacetime) and the richness of the transformation lay  
on the choice of the null vector fields $\vec{\ell}$ used. In this  
sense, we apply the GKS transformation to the FLRW models with constant  
scale factor and taking their most general shear--free geodesic null  
(SFGN hereafter) vector field as the vector field $\vec{\ell}$. These  
objects, which are the basic ingredients to carry out the GKS  
transformation, are studied in section \ref{sec2}. In that situation,  
we study the perfect--fluid Einstein field equations for the GKS  
metrics, from which we obtain the form of the GKS energy--momentum  
tensor (energy density, pressure and fluid velocity) and some partial  
differential equations for the function $\hh$ (section \ref{sec3}).  
Then, through the study of the kinematical quantities of  
the GKS fluid velocity we impose some rectrictions in order to find  
stationary perfect--fluid solutions.  Moreover, these restrictions  
together with the choice of the null basis will allow us to integrate the  
equations for $\hh$ without the explicit use of a coordinate system. In  
addition, we deduce the Petrov type of the GKS spacetimes as well as  
expressions for the non--zero kinematical quantities of the fluid  
velocity (section \ref{sec4}).  
 
In section \ref{sec5}, we find the explicit form of the solutions.  
As far as we know, this family was previously unknown. It is noted that  
it contains all the classical vacuum Kerr--Schild spacetimes, excepting  
the plane--fronted gravitational waves, and hence the Kerr metric  
\cite{KERR} is included too. It is also pointed out that the  
generalization of the Kerr metric in the Einstein background, due to  
Vaidya \cite{VAID}, is also included. However, the family of stationary  
spacetimes found is much more general and depends on an arbitrary  
function of one variable. On the other hand, the efficiency of the  
procedure is shown in section \ref{sec6}, where the Killing  
equations are completely solved without using coordinates. It is  
remarkable that, in general, these spacetimes have only one symmetry.    
Moreover, we discuss the conditions in which other symmetries can  
appear and their consequences on the GKS spacetimes.   
 
The main equations in this paper are written in the Newman--Penrose 
formalism \cite{NEPE}. The conventions that we use here are the 
same of \cite{KSHM} with the only exception that the name 
for the main vector field of the null basis here is $\vec{\ell}$ 
whereas in \cite{KSHM} it is $\vec{\kk}$. Moreover, throughout this 
paper we have used units in which $8\,\pi\,G = c = 1$. Latin indexes 
run from $0$ to $3$. The abbreviation ``c.c." will stand for complex 
conjugate. 
 
\section{Basic ingredients for the GKS transformation} \label{sec2} 
\mbox{} \indent 
In the classical Kerr-Schild transformation the seed spacetime was 
the simplest one, the Minkowski spacetime, being the important object 
the null vector field of the transformation, which was the most  
general SFGN vector field for Minkowski spacetime.  Following this line, 
we take as seeds the subclass of the FLRW spacetimes with constant 
scale factor and their most general SFGN vector field as the vector 
field $\vec{\ell}$ for the transformation.  
 
In what follows, we give the characterization and the explicit form  
of the seed spacetimes $g_{ab}$ as well as of the SFGN vector fields  
$\vec{\ell}$. 
 
\vspace{5mm} 
 
\noindent{\large{\bf The seed metrics}} 
 
\vspace{5mm} 
 
As we have said, the seed spacetimes are the FLRW spacetimes with  
constant scale factor, or equivalently, without expansion.  Taking into  
account the form of the FLRW spacetimes given in appendix A, the line  
element for the seed spacetimes can be written in the following  
explicitly conformally--flat form 
\begin{equation} 
ds^2 = \frac{a^{2}}{(1+\varepsilon\,U^2)(1+\varepsilon\,V^2)}\left 
\{-4\,dU dV + 4\left(\frac{V-U}{1+\xi\bar{\xi}}\right)^{2} d\xi  
d\bar{\xi}\right\}, \label{seed} 
\end{equation} 
where $a$ is the constant scale factor and a bar denotes complex 
conjugation. In the case $\varepsilon = 0$ this line element  
corresponds to the Minkowski spacetime in double null coordinates; in  
the case $\varepsilon = 1$ to the Einstein static Universe and, in  
the case $\varepsilon = -1$, it corresponds to a spacetime usually 
not considered because the energy density is negative, however we  
consider it as a seed spacetime because it may lead to GKS solutions  
with good physical properties. 
 
These conformally--flat spacetimes have a perfect--fluid matter  
content and therefore, its energy--mo\-men\-tum tensor is given by  
\[ T_{ab} = (\varrho+p)\,u_{a}u_{b} + p\,g_{ab} , \hspace{1cm} 
g_{ab}\,u^{a}u^{b} = -1 , \] 
where the energy density $\varrho$ and the pressure $p$ are constant  
\[ \varrho = \frac{3\,\varepsilon}{a^{2}} = -3\,p ,\] 
and in addition, the fluid velocity is 
\[ \vec{u} = u^{a}\,\frac{\partial}{\partial x^{a}} = \frac{\partial} 
{\partial\,t} = \frac{1}{2\,a}\,\left\{ (1+\varepsilon\,U^2)\, 
\frac{\partial}{\partial U} + (1+\varepsilon\,V^2)\,\frac{\partial} 
{\partial V} \right\} ,\]  
\[ {\bf u} = u_{a}\,d x^{a} = -dt = -a\,\left\{ \frac{dU}{1+ 
\varepsilon\,U^2} + \frac{dV}{1+\varepsilon\,V^2} \right\}  . \] 
It can be checked that $\vec{u}$  is a constant vector field, that 
is $\nabla^{}_{\!a}u^{}_{b} = 0$. On the other hand, the case  
$\varepsilon = 0$ is Minkowski ($\varrho=p=0$), and although we can  
define $\vec{u}$, it is not a preferred vector field.  
 
\vspace{5mm} 
 
\noindent{\large{\bf The shear--free geodesic null vector 
field $\vec{\ell}$.}} 
 
\vspace{5mm} 
 
In analogy with the pioneering works \cite{DEKS} on the classical  
Kerr--Schild transformation, we restrict ourselves to the class of  
shear--free geodesic null vector fields.  Although the  
Goldberg--Sachs theorem \cite{GOSA} tells us that such a kind of  
vector fields may not exist in general spacetimes, it is common  
knowledge that in the Minkowski spacetime there is a big family  
of such vector fields and all of them are known explicitly: this result  
constitutes the so--called {\em Kerr's theorem} (see \cite{KSHM,COFL}). 
Moreover, it is well known that a SFGN vector field for Minkowski is  
also a SFGN vector field for any conformally--flat spacetime, 
which means that we have the same large class of such  
vector fields for the seed spacetimes (\ref{seed}).  Here, we are  
going to construct all of them and after this, we will construct an  
appropriate null basis associated with them. 
 
To begin with, the most general null one--form field for the seed  
spacetimes (\ref{seed}) can be written in the following form 
\begin{equation} 
\mbox{\boldmath $\ell$} = F\, \left\{\,dU + Y\,\bar{Y}\,dV +  
\frac{V-U}{1+\xi\bar{\xi}}\left(\bar{Y}\,d\xi + Y\,d\bar{\xi}\, 
\right)\, \right\} , \label{sgnv} 
\end{equation} 
where $F$ and $Y$ are real and complex arbitrary scalar fields 
respectively. Then, in order that {\boldmath $\ell$} be 
geodesic and shear--free,  the complex function $Y$ must 
be a solution of the following system of non--linear partial  
differential equations 
\begin{equation} 
(1+\xi\bar{\xi})\,Y\,Y_{,\xi} = (V-U)\,Y_{,V} + Y\,(1+\bar{\xi}\,Y)  
\, , \label{yec1} 
\end{equation} 
\begin{equation} 
(1+\xi\bar{\xi})\,Y_{,\bar{\xi}} = (V-U)\,Y\,Y_{,U} - Y\,(Y-\xi)  
\, , \label{yec2} 
\end{equation} 
where commas stand for partial derivative with respect to the  
subscript that follows.   
 
Now, in order to simplify further calculations, we take a null basis  
$\{ \ell, \kk, \mm, \bar{\mm} \}$ for the seed spacetimes (\ref{seed})  
associated with $\vec{\ell}$.  We choose this null basis in such a way  
that $\vec{\ell}$ be affinely para\-me\-tri\-zed and such that the  
unit timelike vector field $\vec{u}$ can be written in the following  
way 
\[ \vec{u} = \frac{1}{\sqrt{2}}\left(\vec{\ell} + \vec{\kk}\,\right), 
\hspace{1cm} {\bf u} = \frac{1}{\sqrt{2}}\left( \mbox{\boldmath $\ell$} 
 + {\bf \kk} \right) \, . \] 
Part of the remaining freedom is used for setting the imaginary part  
of the spin coefficient $\epsilon$ equals to zero.  After some  
calculations, the explicit expressions for a such null basis  
$\{\ell,\kk,\mm,\bar{\mm}\}$ are (\ref{sgnv}) and 
\[ {\bf \kk} = F\, \left\{\,\frac{1+\varepsilon\,V^2}{1+ 
\varepsilon\,U^2}\, Y\,\bar{Y}\,dU+\frac{1+\varepsilon\,U^2}{1+ 
\varepsilon\,V^2}\,dV - \frac{V-U}{1+\xi\bar{\xi}} \left(\bar{Y}\,d\xi  
+ Y\,d\bar{\xi}\,\right)\, \right\} \, , \] 
\begin{eqnarray*}  
{\bf \mm} = &-&F\, \left\{\,\sqrt{Y\,\bar{Y}}\left(-\sqrt{ 
\frac{1+\varepsilon\,V^2}{1+\varepsilon\,U^2}}\,dU+\sqrt{\frac{1+ 
\varepsilon\,U^2}{1+\varepsilon\,V^2}}\,dV \right)  \right. + \\ 
& & \left. \frac{V-U}{1+\xi\bar{\xi}} \sqrt{\frac{\bar{Y}}{Y}}\left( 
\sqrt{\frac{1+\varepsilon\,U^2}{1+\varepsilon\,V^2}}\,d\xi - Y^2 
\sqrt{\frac{1+\varepsilon\,V^2}{1+\varepsilon\,U^2}}\,d\bar{\xi}\, 
\right) \, \right\} \, , 
\end{eqnarray*} 
where the case $Y=0$ can be obtained by taking $Y/\bar{Y}=1$, and 
$F$ is the following function  
\[ F = \frac{-\,\sqrt{2}\,a}{(1 + \varepsilon\,U^2)+ (1 + 
\varepsilon\,V^2) Y \bar{Y}} \, . \] 
 
After long but straightforward calculations we find that the spin  
coefficients associated with this null basis satisfy the relations 
\begin{equation} \begin{array}{c} 
\kappa \hspace{3mm} = \hspace{3mm} \sigma \hspace{3mm} =  
\hspace{3mm} \epsilon \hspace{3mm} = \hspace{3mm} \pi \hspace{3mm} 
= \hspace{3mm} \lambda \hspace{3mm} = \hspace{3mm} 0 \, , \\ 
\tau - \bar{\nu} \hspace{3mm} = \hspace{3mm} \rho - \bar{\mu} 
\hspace{3mm} = \hspace{3mm}  \alpha + \bar{\beta} \hspace{3mm} = 
\hspace{3mm} \gamma + \bar{\gamma} \hspace{3mm} = \hspace{3mm} 0  
\, . \end{array} \label{coes} 
\end{equation} 
Furthermore, the Newman--Penrose symbols for the Riemann tensor are 
 
\vspace{5mm} 
 
\noindent{\em Ricci tensor:} 
\[ \Phi_{01}  = \Phi_{02}  = \Phi_{12} = 0, \hspace{2mm}  
\Phi_{00} = 2\,\Phi_{11} = \Phi_{22} = \frac{1}{4}(\varrho + p) 
= \frac{\varepsilon}{2 a^2}  , \hspace{2mm} \Lambda =  
\frac{1}{24}(\varrho-3p) = \frac{\varepsilon}{4 a^2} \, , \] 
 
\vspace{5mm} 
 
\noindent{\em Weyl tensor:} 
\[ \Psi_{0} = \Psi_{1} = \Psi_{2} = \Psi_{3} = \Psi_{4} = 0 \, . \] 
 
\section{The construction of the GKS spacetimes} \label{sec3} 
\mbox{} \indent 
In this section we study the GKS transformation (\ref{gkst}) for the 
seed spacetimes (\ref{seed}) with (\ref{sgnv}) as the SFGN vector  
field $\vec{\ell}$. To that end, we use the fact that given a null  
basis $\{ \ell, \kk, \mm, \bar{\mm} \}$ associated with the seed 
metric, we can construct a null basis $\{ \tilde{\ell}, \tilde{\kk}, 
\tilde{\mm}, \tilde{\bar{\mm}} \}$ associated with the GKS metric by 
taking (see \cite{TAUB,MASE}) 
\begin{equation} 
\begin{array}{lcl} 
\tilde{\ell}^{a} \hspace{2mm} = \hspace{2mm} \ell^{a} \, ,  
& \hspace{2cm} & 
\tilde{\ell}_{a} \hspace{2mm} = \hspace{2mm} \ell_{a} \, , \\ 
\tilde{\kk}{}^{a} \hspace{2mm} = \hspace{2mm} \kk^{a} + \hh\,\ell^{a}  
\, , & \hspace{2cm} &   
\tilde{\kk}_{a} \hspace{2mm} = \hspace{2mm} \kk_{a} - \hh\,\ell_{a}  
\, , \\  
\tilde{\mm}^{a} \hspace{2mm} = \hspace{2mm} \mm^{a} \, ,  
& \hspace{2cm} & 
\tilde{\mm}_{a} \hspace{2mm} = \hspace{2mm} \mm_{a} \, , 
\end{array}  \label{cht1} 
\end{equation} 
where, from now on, objects with tilde are associated with the GKS  
spacetimes in contrast with objects without tilde which are associated  
with the seed spacetimes. 
 
These relationships between null bases allow us to compute all objects  
associated with the GKS spacetimes as functions of the same objects  
associated with the seed spacetimes, and the function $\hh$ and their  
derivatives.  In this sense, we can compute all the Newman--Penrose 
symbols for the GKS spacetimes.  Taking into account the 
expressions of the previous section, the result for the components 
of the Ricci tensor is the following 
\begin{equation} 
\tilde{\Phi}_{00} = \frac{\varepsilon}{2\,a^{2}} , \hspace{15mm} 
\tilde{\Phi}_{01} = \tilde{\Phi}_{02} =  0 , \label{ph00} 
\end{equation} 
\begin{equation} 
\tilde{\Phi}_{11} = \frac{\varepsilon}{4\,a^{2}} + \frac{1}{2}\left[ 
(\rho-\bar{\rho})^2 + \frac{\varepsilon}{2\,a^{2}} \right]\,\hh +  
\frac{1}{4}(\rho+\bar{\rho})\,\vv - \frac{1}{4} D\,\vv \, , 
\end{equation} 
\begin{equation}  
\tilde{\Phi}_{12} = \left[ \delta \bar{\rho} + (\rho-\bar{\rho})  
\tau \right]\,\hh + \bar{\rho}\,\hh + \frac{1}{2}\left(\tau\, 
\vv - \delta\,\vv \right) \, , 
\end{equation} 
\begin{eqnarray} 
\tilde{\Phi}_{22} &=& \frac{\varepsilon}{2\,a^{2}} + \frac{1}{2} 
\left[2\,\triangle (\rho + \bar{\rho}) - (\rho-\bar{\rho})^2 \right] 
\,\hh + \frac{1}{2}(\rho+\bar{\rho})\,\triangle\,\hh + \frac{1}{2} 
(\rho+\bar{\rho})\,\vv +  \nonumber \\ 
& & (\bar{\tau}+\alpha)\,\uu + (\tau+\bar{\alpha})\,\bar{\uu} - 
\frac{1}{2}\left(\delta\,\bar{\uu} + \bar{\delta}\,\uu \right) + 
\frac{\varepsilon}{2\,a^{2}}\,\hh^2 \, , 
\end{eqnarray} 
\begin{equation} 
\tilde{\Lambda} = \frac{\varepsilon}{4\,a^{2}} + \frac{1}{6} 
\left[(\rho+\bar{\rho})^2 +2\,\rho\bar{\rho} - \frac{3\, 
\varepsilon}{2\,a^{2}}\right]\,\hh - \frac{1}{4}(\rho+\bar{\rho}) 
\,\vv + \frac{1}{12} D\,\vv \, , \label{lamb} 
\end{equation} 
where $\vv$ and $\uu$ are defined by 
\[ \vv \equiv D\,\hh + (\rho+\bar{\rho})\,\hh , \hspace{15mm} \uu 
\equiv \delta\,\hh \, . \] 
 
In the same way, we can obtain the components of the Weyl tensor.  
Their expressions are 
\begin{equation} \tilde{\Psi}_{0} = 0 \, , \hspace{5mm} 
\tilde{\Psi}_{1} = 0 \, , \label{psi0} 
\end{equation} 
\begin{equation} 
3\,\tilde{\Psi}_{2} = -(\rho-\bar{\rho})^2\,\hh - \frac{3}{2} 
(\rho-\bar{\rho})\,\vv - \frac{1}{2} D\,\vv \, , \label{psi2} 
\end{equation} 
\begin{equation} 
\tilde{\Psi}_{3} = -(\rho-\bar{\rho})\,\bar{\uu} + \frac{1}{2} 
(\bar{\tau}\,\vv - \bar{\delta}\,\vv) \, , \hspace{5mm} 
\tilde{\Psi}_{4} = 2\,(\bar{\tau} - \alpha)\,\bar{\uu} - 
\bar{\delta}\,\bar{\uu} \, . \label{psi4} 
\end{equation} 
As we can see directly from these equations, all the GKS spacetimes  
that we can obtain are algebraically special because the null vector  
field $\vec{\tilde{\ell}} = \vec{\ell}$ is a multiple null  
eigenvector of the Weyl tensor. 
 
Now, we must study the Einstein field equations for the GKS 
metrics with a perfect--fluid source. They read as follows 
\begin{equation} 
\tilde{G}_{ab} + C \tilde{g}_{ab} \hspace{3mm} = \hspace{3mm}  
\tilde{T}_{ab} \, , \label{tein} 
\end{equation} 
where $\tilde{G}_{ab}$ is the Einstein tensor for the GKS metrics,  
$C$ is the cosmological constant, and 
$\tilde{T}_{ab}$ is the energy--momentum tensor, which has the form 
\begin{equation} 
\tilde{T}_{ab} = (\tilde{\varrho}+\tilde{p})\,\tilde{u}_{a} 
\tilde{u}_{b} + \tilde{p}\,\tilde{g}_{ab} , \hspace{1cm}  
\tilde{g}_{ab}\,\tilde{u}^{a} \tilde{u}^{b} = -1 \, , \label{ttem} 
\end{equation} 
where $\tilde{\varrho}$, $\tilde{p}$ and $\vec{\tilde{u}}$ are the  
energy density, the pressure and the fluid velocity of the GKS  
perfect fluid, respectively. In this situation, we project the  
Einstein equations (\ref{tein}) onto the null basis  
$\{\tilde{\ell},\tilde{\kk},\tilde{\mm},\tilde{\bar{\mm}}\}$ and  
then, using (\ref{ttem}) we obtain other expressions for the Ricci  
tensor components 
\begin{equation} 
\tilde{\Phi}_{00} = \frac{1}{2}\left(\tilde{\varrho}+\tilde{p} 
\right)(\tilde{\ell}^{a}\,\tilde{u}_{a})^{2} \, , \label{tf00} 
\end{equation} 
\begin{equation} 
\tilde{\Phi}_{01} = \frac{1}{2}\left(\tilde{\varrho}+\tilde{p} 
\right)(\tilde{\ell}^{a}\,\tilde{u}_{a})\,(\tilde{\mm}^{b}\, 
\tilde{u}_{b}) \, , \hspace{10mm} \tilde{\Phi}_{02} = \frac{1}{2} 
\left(\tilde{\varrho}+\tilde{p}\right) (\tilde{\mm}^{a}\, 
\tilde{u}_{a})^{2} \, , \label{tf02} 
\end{equation} 
\begin{equation} 
\tilde{\Phi}_{11} = \frac{1}{4}\left(\tilde{\varrho}+\tilde{p} 
\right)\left[(\tilde{\ell}^{a}\,\tilde{u}_{a})\, 
(\tilde{\kk}{}^{b}\,\tilde{u}_{b})+(\tilde{\mm}^{a}\, 
\tilde{u}_{a})\,(\tilde{\bar{\mm}}^{b}\,\tilde{u}_{b}) \right]  
\, , \label{tf11} 
\end{equation} 
\begin{equation} 
\tilde{\Phi}_{12} = \frac{1}{2}\left(\tilde{\varrho}+\tilde{p} 
\right)(\tilde{\kk}{}^{a}\,\tilde{u}_{a})\,(\tilde{\mm}^{a}\, 
\tilde{u}_{a}) \, , \hspace{10mm} \tilde{\Phi}_{22} = \frac{1}{2} 
\left(\tilde{\varrho}+\tilde{p}\right)(\tilde{\kk}{}^{a}\, 
\tilde{u}_{a})^{2} \, , \label{tf22} 
\end{equation} 
\begin{equation} 
\tilde{\Lambda} = \frac{1}{24}\left(\tilde{\varrho} - 3\,\tilde{p} 
\right)+\frac{1}{6}C \, . \label{tlam} 
\end{equation} 
 
The next step in this process is to compare the expressions for the  
Ricci tensor (\ref{ph00}--\ref{lamb}) with the Einstein equations  
for the GKS metrics (\ref{tf00}--\ref{tlam}).  We do this for the  
seed metrics with $\varepsilon \neq 0$, so that $\tilde{\varrho} + 
\tilde{p} \neq 0$,  and later we will extend the results for the 
$\varepsilon = 0$ case.  The outcome of this comparison is 
 
\vspace{5mm} 
 
$\bullet$ The components, in the null bases $\{\ell,\kk,\mm,\bar{\mm} 
\}$ and $\{\tilde{\ell},\tilde{\kk},\tilde{\mm},\tilde{\bar{\mm}}\}$, 
for the fluid velocity $\vec{\tilde{u}}$ of the GKS perfect fluid: 
\begin{equation} 
(\tilde{\ell}^{a}\,\tilde{u}_{a})^2 = (\ell^{a}\,\tilde{u}_{a})^2 = 
\frac{\varepsilon}{2 a^2}\,\left[ \frac{\varepsilon}{a^2} + 2\,\left( 
(\rho-\bar{\rho})^2+ \frac{\varepsilon}{2 a^2} \right)\, 
\hh + (\rho+\bar{\rho})\,\vv - D\,\vv \right]^{-1} \, , \label{utl} 
\end{equation} 
\begin{equation} \tilde{\kk}{}^{a}\,\tilde{u}_{a} = \kk^{a}\, 
\tilde{u}_{a} + \hh\,\ell^{a}\,\tilde{u}_{a} = \left[2\, 
(\tilde{\ell}^{a}\,\tilde{u}_{a}) \right]^{-1} \, , \label{utk} 
\end{equation} 
\begin{equation} 
\tilde{\mm}^{a}\,\tilde{u}_{a} = \mm^{a}\,\tilde{u}_{a} = 0  
\, . \label{utm} 
\end{equation} 
 
\vspace{5mm} 
 
$\bullet$ The energy density and the pressure of the GKS perfect  
fluid: 
\begin{equation} 
\tilde{\varrho} = \frac{3\,\varepsilon}{a^2}-C+2\,(2\,\rho^2-\rho\, 
\bar{\rho}+2\,\bar{\rho}^2)\,\hh - D\,\vv \, , \label{roks} 
\end{equation} 
\begin{equation} 
\tilde{p} = -\frac{\varepsilon}{a^2}+C+2\,\left(-3\,\rho\,\bar{\rho}+ 
\frac{\varepsilon}{a^2} \right)\,\hh + 2\,(\rho+\bar{\rho})\,\vv 
- D\,\vv \, , \label{prks} 
\end{equation} 
 
\vspace{5mm} 
 
$\bullet$ Two second order partial differential equations for $\hh$: 
\begin{equation} 
\delta\,\vv = \tau\,\vv + 2\,\bar{\rho}\,\uu + 2\,\left[\delta\, 
\bar{\rho} + (\rho-\bar{\rho})\,\tau\right]\,\hh \, , \label{ec01} 
\end{equation} 
\begin{eqnarray} 
(\rho+\bar{\rho})\,\triangle\,\hh & = &\left[(\rho-\bar{\rho})^2 + 
\frac{\varepsilon}{a^2} -2\,\triangle\,(\rho+\bar{\rho})\right]\,\hh - 
(\rho+\bar{\rho})\,\vv +  \nonumber \\ 
& &\delta\,\bar{\uu} + \bar{\delta}\,\uu - 
2\,(\bar{\tau}+\alpha)\,\uu - 2\,(\tau+\bar{\alpha})\,\bar{\uu} + 
(\frac{\varepsilon}{a^2})^{-1}\,\ww \, ,  \label{ec02} 
\end{eqnarray} 
where $\ww$ is given by 
\begin{eqnarray} 
&\ww& = \left[D\,\vv-(\rho+\bar{\rho})\,\vv-2(\rho-\bar{\rho})^2\, 
\hh\,\right]\,\times  \nonumber \\ 
& & \left[D\,\vv-(\rho+\bar{\rho})\,\vv-2(\rho-\bar{\rho})^2\,\hh 
- \frac{2\,\varepsilon}{a^2}\,(1+\hh) \right] \, . \label{wwww} 
\end{eqnarray} 
As we can see, equation (\ref{ec02}) is a non--linear partial 
differential equation for $\hh$ due to the term with $\ww$. 
 
\section{Characterization of the GKS space\-times through the  
kinematical quantities} \label{sec4} 
\mbox{}\indent 
Until now, the only assumption we have made on the GKS spacetimes is  
that their energy--momentum tensor must be of the perfect--fluid type.   
Then, once we have a function $\hh$ solution of the system  
(\ref{ec01}--\ref{ec02}), we have the explicit form of the GKS  
metric as well as of the matter content variables.  However, there  
is not a systematic way of solving (\ref{ec01}--\ref{ec02}),  
specially because (\ref{ec02}) is a non--linear equation.  For this  
reason, we are going to introduce some additional assumptions on the  
GKS spacetimes which can help us to integrate these equations. Of  
course, it would be advisable to impose conditions which have a  
physical meaning. In this sense, an interesting way to control these  
conditions is through the study of the kinematical quantities for the  
fluid velocity $\vec{\tilde{u}}$ of the GKS perfect fluid.  
 
From equations (\ref{utl}--\ref{utm}) it follows that  
$\vec{\tilde{u}}$ lies in the two--planes generated by  
$\vec{\tilde{\ell}}$ and $\vec{\tilde{\kk}}$. Then, by making the  
following change of basis 
\begin{equation} 
\vec{\tilde{\ell}} \rightarrow \vec{L} = \aaa\,\vec{\tilde{\ell}}  
\, , \hspace{1cm} \vec{\tilde{\kk}} \rightarrow \vec{K} = \aaa^{-1} 
\vec{\tilde{\kk}} \, , \hspace{1cm} \vec{\tilde{\mm}} \rightarrow  
\vec{\tilde{\mm}} \, , \label{cht2} 
\end{equation} 
where 
\begin{equation} 
\aaa^2 \equiv \left[2\,(\ell^{a}\tilde{u}_{a})^2\right]^{-1}  
\, , \label{adef} 
\end{equation} 
we get the following form for $\vec{\tilde{u}}$ 
\[ \vec{\tilde{u}} \hspace{3mm} = \hspace{3mm} \frac{1}{\sqrt{2}}  
(\vec{L} + \vec{K} ) \, . \] 
Therefore, we can now use the formulas given in the appendix  
\ref{appb} to obtain the kinematical quantities of $\vec{\tilde{u}}$.  
To that end, we compute the spin coefficients associated with the  
null basis $\{ L, K, \tilde{\mm}, \tilde{\bar{\mm}}\}$, which we  
denote with a hat. The most useful results are 
\[ \hat{\rho}+\hat{\bar{\rho}} = (\rho+\bar{\rho})\,\aaa \, ,  
\hspace{1cm} \hat{\mu}+\hat{\bar{\mu}} = (\rho+\bar{\rho})(1-\hh) 
\,\aaa^{-1} \, , \hspace{1cm} \hat{\epsilon}+\hat{\bar{\epsilon}} =  
D\aaa \, , \] 
\[ \hat{\gamma}+\hat{\bar{\gamma}} = -\aaa^{-1} D\hh + (\triangle\aaa  
+\hh\,D\aaa)\,\aaa^{-2} \, , \hspace{1cm} \hat{\alpha}+ 
\hat{\bar{\beta}} = \aaa^{-1} \bar{\delta}\aaa \, , \] 
\[ \hat{\pi}+\hat{\bar{\tau}}-\hat{\nu}-\hat{\bar{\kappa}} = 
\bar{\tau} + \left[\,\bar{\delta}\,\hh - \bar{\tau}(1+\hh)\right]\, 
\aaa^{-2} , \hspace{1cm} \hat{\sigma} = \hat{\lambda} = 0 \, , \] 
\[ \hat{\rho}-\hat{\bar{\rho}} = (\rho-\bar{\rho})\,\aaa \, ,  
\hspace{1cm} \hat{\mu}-\hat{\bar{\mu}} = -(\rho-\bar{\rho})(1+\hh)\, 
\aaa^{-1} \, , \] 
 
In this situation, we are going to impose conditions which can lead  
to stationary spacetimes. To that end, we impose the vanishing of the  
shear and expansion of the fluid velocity $\vec{\tilde{u}}$, that is, 
\begin{equation} 
\tilde{\theta} \hspace{3mm} = \hspace{3mm} \tilde{\sigma}_{ab} 
\hspace{3mm} = \hspace{3mm} 0 \, . \label{exsh} 
\end{equation} 
These conditions are necessary in order to have $\vec{\tilde{u}}$ 
proportional to a time--like Killing vector field. Then, from the 
vanishing of the expansion and the $\mm_{(a}\bar{\mm}_{b)}$--component 
of the shear tensor (see appendix \ref{appb}) we have 
\[ (\rho+\bar{\rho})\left( \aaa^2 - 1 + \hh \right)  = 0, \] 
so that, two different cases may appear: $\rho+\bar{\rho} 
= 0 $ or $\aaa^2 = 1 - \hh$. If $\rho+\bar{\rho} = 0$ the SFGN 
vector field $\vec{\ell}$ is expansion--free. In addition, from the  
Newman--Penrose equations \cite{NEPE} we obtain that $\rho^{2}+ 
\varepsilon/(2\,a^2) = 0$, which means that only the case  
$\varepsilon=1$ is possible. However, further analysis of the  
equations for this case shows that there are not solutions in this 
case.  Therefore, from now on we will consider that  
$\rho+\bar{\rho}\neq 0$, and then we must follow through the second  
possibility  
\begin{equation} 
\aaa^{2} \hspace{2mm} = \hspace{2mm} 1 - \hh \, .  \label{aexp} 
\end{equation} 
From this equation and using the expressions (\ref{utl}, 
\ref{adef}) we get another differential equation for $\hh$ 
\begin{equation} 
D\,\vv = (\rho+\bar{\rho})\,\vv + 2\left[ (\rho-\bar{\rho})^2 + 
\frac{\varepsilon}{a^2} \right]\,\hh  \, . \label{ec03} 
\end{equation} 
If we use this equation, we can see from (\ref{wwww}) that 
$\ww=-4\,(\varepsilon^{2}/a^2)\,\hh$, and therefore  
(\ref{ec02}) is a linear differential equation for $\hh$.  Moreover, 
from the remaining components of equations (\ref{exsh}) we have 
the following two additional conditions on $\hh$ and $Y$, 
\begin{equation} 
 (D+\triangle)\,\hh = 0 \, , \label{ec04} 
\end{equation} 
\begin{equation} 
\tau = 0 \, , \label{tau0} 
\end{equation} 
respectively.  
The first is a further differential equation for $\hh$, while the  
second one is a constraint on the possible complex functions $Y$  
that we can use to construct the SFGN vector fields $\vec{\ell}$. Then,  
replacing (\ref{aexp}) into expressions (\ref{utl}--\ref{utm}) for  
the fluid velocity $\vec{\tilde{u}}$ we obtain 
\begin{equation} 
\vec{\tilde{u}} = \frac{1}{\sqrt{2\,(1-\hh)}}\left[(1-\hh)\, 
\vec{\tilde{\ell}}+\vec{\tilde{\kk}}\,\right] = \frac{1}{\sqrt{2\, 
(1-\hh)}}\left(\vec{\ell}+\vec{\kk}\,\right) = \frac{1} 
{\sqrt{1-\hh}}\,\vec{u} \, , \label{vecu} 
\end{equation} 
that is to say, $\vec{\tilde{u}}$ is proportional to $\vec{u}$. On  
the other hand, it can be shown that the vector field $\vec{u}$,  
which is a Killing vector field for the seed spacetimes (\ref{seed}),  
is also a Killing vector field for the GKS spacetimes provided that  
(\ref{ec04}) and (\ref{tau0}) hold.  Thus, $\vec{\tilde{u}}$ is  
parallel to a Killing vector field. We can see this fact from the  
following expression  
\begin{equation} 
\tilde{\nabla}^{}_{\!a}\tilde{u}'{}^{}_{b} +  
\tilde{\nabla}^{}_{\!b}\tilde{u}'{}^{}_{a} = \sqrt{2}\left\{ 
( D + \triangle ) \hh \ell^{}_{a}\ell^{}_{b} +  
2\left[ \tau \kk^{}_{(a}\mm^{}_{b)} + \mbox{c.c.} \right]\hh 
\right\} = 0 \, , 
\end{equation} 
where $\tilde{u}'{}^{}_{a} = \tilde{g}^{}_{ab}u^{b}$. Here, we have 
used (\ref{coes}) and the second equality follows from  
(\ref{ec04},\ref{tau0}).  Notice that this is not a trivial result  
because, even though conditions (\ref{exsh}) are necessary for  
$\vec{\tilde{u}}$ to be proportional to a Killing, they are not 
sufficient (we also need the acceleration ${\bf \tilde{\acc}}$  
to be an exact 1--form).  Therefore, we have proved that (\ref{exsh}) implies  
${\bf d\tilde{\acc}} = {\bf 0}$ under our conditions.

On the other hand, in the last section we restricted ourselves to  
the cases $\varepsilon = \pm 1$.  Now, we can include the case 
$\varepsilon = 0$ by defining the vector field $\vec{\tilde{u}}$  
through the expression (\ref{vecu}), and by taking the same partial 
differential equations (\ref{ec01}--\ref{ec02}) and 
(\ref{ec03}--\ref{ec04}) for $\hh$. 
 
The next point is to find which are the integrability conditions  
for the system of \underline{linear} partial differential equations 
for $\hh$. In this process we find two different cases. 
 
\vspace{5mm} 
 
$\bullet$ \underline{Case A}:\hspace{3mm}$\bar{\delta} \rho = 0 
\hspace{3mm} \Rightarrow \rho - \bar{\rho} = 0$. It can be shown 
that all the GKS metrics that we can find in this case are included 
in a particular case of reference \cite{MPSE}, where the authors use 
the GKS transformation with the interior Schwarzschild metric as the 
seed metric.  The resultant GKS spacetimes in that paper are  
stationary spherically symmetric and they contain a subfamily of  
regular solutions with equation of state $\tilde{\varrho}+3\,\tilde{p}  
= \mbox{const.}$. Among these spacetimes we have the Whittaker metric  
\cite{WHIT} as a special case, which, at the same time, is the static  
limit of the Wahlquist metric \cite{WAHL}.  
 
\vspace{5mm} 
 
$\bullet$ \underline{Case B}:\hspace{3mm}$\bar{\delta} \rho \neq 0$. 
In this case the procedure to find the integrability conditions of the 
system of partial differential equations (\ref{ec01}--\ref{ec02}) and 
(\ref{ec03}--\ref{ec04}) for $\hh$, leads us to the following  
equations for the first derivatives of $\hh$ 
\begin{equation} 
D\,\hh = \frac{\rho^2+\bar{\rho}^2+ \frac{\varepsilon}{a^2}}{\rho+ 
\bar{\rho}}\,\hh = - \triangle\,\hh , \hspace{1cm} \delta\,\hh =  
\frac{\delta \bar{\rho}}{\rho+\bar{\rho}}\,\hh \, . \label{dthh} 
\end{equation} 
These equations can be easily solved without the explicit use of  
coordinates and the solution is 
\begin{equation} 
\hh \hspace{2mm} = \hspace{2mm} m\,(\rho+\bar{\rho})\, , \label{hhhh} 
\end{equation} 
where $m$ is an arbitrary constant. 
 
The energy--momentum tensor of these GKS spacetimes can be obtained  
by substituting $\hh$ in the explicit form of the fluid velocity,  
the energy density and the pressure, given in (\ref{vecu}) and  
(\ref{roks}--\ref{prks}), respectively. In particular, the energy  
density and the pressure are 
\begin{equation} 
\tilde{\varrho} = \frac{3\,\varepsilon}{a^2}\,\left[1-m\,(\rho+ 
\bar{\rho})\right]-C \, , \hspace{5mm} \tilde{p} = \frac{\varepsilon} 
{a^2}\, \left[-1+m\,(\rho+\bar{\rho})\right]+C \, , \label{rpgk}  
\end{equation} 
so that we have the following equation of state 
\[ \tilde{\varrho} + 3\,\tilde{p} = 2 C \, . \] 
The most interesting case from the physical point of view is the one 
given by $C\geq 0$ and $\varepsilon=1$, for which there are spacetime  
regions where all the energy conditions are fulfilled  
(see \cite{HAEL}).  In this case, from the form (\ref{vecu}) of  
$\vec{\tilde{u}}$, it follows that the region of the spacetime in  
which we can define this unit timelike vector field is 
\[ 1 - \hh \hspace{3mm} > \hspace{3mm} 0 \, . \] 
Moreover, the case $\varepsilon=C=0$ corresponds to vacuum solutions.  
In the next section we will analyze to which solutions correspond  
each case.

On the other hand, from the expressions (\ref{psi0}--\ref{psi4}) for  
the Weyl tensor we already know that the GKS spacetimes are  
algebraically special.  In this case, the Petrov type is II in  
general, except when the following relation holds 
\begin{equation} 
3\,\tilde{\Psi}_{2}\tilde{\Psi}_{4} = 2\,\tilde{\Psi}^{2}_{3} 
\hspace{3mm} \Longrightarrow \hspace{3mm} (\,\rho^2+\frac{\varepsilon} 
{2\,a^2})\,(\,\bar{\delta}\bar{\delta} \rho + 2\,\alpha\,\bar{\delta} 
\rho ) = 3\,\rho\,(\bar{\delta} \rho )^2 \, . \label{petr} 
\end{equation} 
In that case the Petrov type is D and the fluid velocity does not  
lie in the preferred two--space spanned by the two multiple null 
eigenvectors of the Weyl tensor. 
 
Finally, the non vanishing kinematical quantities for  
$\vec{\tilde{u}}$ are 
\[ \vec{\tilde{\mbox{a}}} = -\frac{D\,\hh}{\sqrt{2\,(1-\hh)}} 
\vec{\tilde{v}}-\left(\frac{\bar{\delta}\,\hh}{1-\hh} 
\vec{\tilde{\mm}}+\mbox{c.c.}\right) \, , \] 
\begin{equation} 
\tilde{\omega}_{ab} = - \left(\frac{\bar{\delta}\,\hh}{1-\hh}  
\tilde{v}_{[a}\tilde{\mm}_{b]} + \mbox{c.c.} \right) -  
\frac{\sqrt{2}\,(\rho-\bar{\rho})\,\hh}{\sqrt{1-\hh}}\, 
\tilde{\mm}_{[a}\,\tilde{\bar{\mm}}_{b]} \, , \label{rott} 
\end{equation} 
where $\tilde{v}^{a}$ is a unit spacelike vector field, orthogonal  
to $\tilde{u}^{a}$, and defined as follows 
\[ \tilde{v}^{a} \equiv \frac{(1-\hh)\tilde{\ell}^{a}- 
\tilde{\kk}^{a}}{\sqrt{2\,(1-\hh)}} \hspace{4mm} \Longrightarrow 
\hspace{4mm} \tilde{v}^{a}\tilde{v}_{a} = 1 \, , \hspace{2mm} 
\tilde{v}^{a}\tilde{u}_{a} = 0 \, . \] 
It is interesting to note that from expression (\ref{rott}) and  
from equations (\ref{dthh}-\ref{hhhh}) we can deduce that the  
rotation $\tilde{\omega}_{ab}$ of \, $\tilde{u}^{a}$ vanishes if  
and only if so does the rotation $\rho-\bar{\rho}$ of $\ell^{a}$. 
 
\section{Explicit form of the solutions} \label{sec5} 
\mbox{}\indent 
After the integration of the equations for $\hh$, we are going to  
find an explicit expression for the line element of the GKS  
spacetimes. For this we must find explicitly all the SFGN vector  
fields $\vec{\ell}$ satisfying the condition (\ref{tau0}), that is,  
we must find the explicit solutions of the partial differential  
equations (\ref{yec1}--\ref{yec2}) together with (\ref{tau0}). To do  
that, it is more convenient to consider a new function $\Omega$  
instead of $Y$, which is defined by 
\[ Y(U,V,\xi,\bar{\xi}) \equiv \sqrt{\frac{1+\varepsilon\,U^2}{1+ 
\varepsilon\,V^2}}\,\Omega(t,\chi,\xi,\bar{\xi}) , \] 
and also, it is better to change from the coordinates $\{U,V\}$ to  
the $\{t,\chi\}$ ones (see appendix \ref{appa}). Once we have  
performed these changes, the equations (\ref{yec1}--\ref{yec2})  
and (\ref{tau0}) read as follows 
\begin{equation} 
(1+\xi\bar{\xi})\,\Omega\,\Omega_{,\xi} - \bar{\xi}\,\Omega^{2} = 
\Sigma\,\Omega_{,\chi} + \Sigma'\,\Omega , \label{oec1} 
\end{equation} 
\begin{equation} 
(1+\xi\bar{\xi})\,\Omega_{,\bar{\xi}} + \xi\,\Omega = -\Sigma\, 
\Omega\,\Omega_{,\chi} + \Sigma'\,\Omega^{2} , \label{oec2} 
\end{equation} 
\begin{equation} 
\Omega_{,t} = 0 , \label{oec3} 
\end{equation} 
where $\Sigma(\varepsilon,\chi)$ is given in appendix A. The most 
general solution of these equations can be given implicitly by 
\begin{equation} 
G\left( \frac{\mbox{e}^{-\sqrt{-\varepsilon\,\chi}}\,\Omega - \xi} 
{1 + \bar{\xi}\,\mbox{e}^{-\sqrt{-\varepsilon\,\chi}}\,\Omega}\, ,  
\, \frac{(1-\xi\bar{\xi})\,\Sigma'\,\Omega - \xi + \bar{\xi}\, 
\Omega^{2}}{(1+\xi\bar{\xi})\,\Sigma\,\Omega} \right) = 0 \,  
, \label{fung} 
\end{equation} 
where $G(z_{1},z_{2})$ is any analytic complex function of two  
complex variables $\{z_{1},z_{2}\}$. It is interesting to note that,  
in the same way, we could have found the general solution of only  
equations (\ref{oec1}) and (\ref{oec2}), then we would have obtained  
all the possible SFGN vector fields for the seed metrics  
(\ref{seed}), what constitutes the analogous result of the Kerr  
theorem.  However, for the sake of brevity we do not give here the  
explicit form of the general solution. From the definition of the  
spin coefficient $\rho$, and using (\ref{oec1}--\ref{oec3}) we have 
\[ \rho = \left\{ \begin{array}{ll} \frac{1}{\sqrt{2}a} 
\frac{\Omega_{,\chi}}{\Omega} & \mbox{for}\hspace{3mm}\Omega\neq 0  
\, , \\ \mbox{} & \mbox{} \\ \frac{1}{\sqrt{2}a}\frac{\Sigma'} 
{\Sigma} &\mbox{for}\hspace{3mm}\Omega=0 \, , \end{array} \right. \] 
and then, the solution (\ref{hhhh}) for $\hh$ is 
\[ \hh = \left\{ \begin{array}{ll} \frac{m}{\sqrt{2}\,a}\, 
\left[\ln(\Omega\, \bar{\Omega})\right]_{,\chi} & \mbox{for} 
\hspace{3mm}\Omega\neq 0 \, , \\ 
\mbox{} & \mbox{} \\ \frac{\sqrt{2}m}{a}\frac{\Sigma'}{\Sigma}  
&\mbox{for} \hspace{3mm}\Omega=0 \, . \end{array} \right.\] 
Now, making the substitution of the expressions (\ref{hhhh}) for the  
function $\hh$ and (\ref{vecu}) for the SFGN 1--form  
{\boldmath $\ell$} into the definition of the GKS transformation  
(\ref{seed}), we find the following explicit form for the line  
element of the GKS metrics  
\begin{eqnarray} 
&&ds^2 = -dt^2+a^2\left(d\chi^2+\frac{4\,\Sigma^2\,d\xi d\bar{\xi}} 
{(1+\xi\bar{\xi})^{2}}\, \right) +  \nonumber \\ 
& & \frac{a^{2}\hh}{(1+\Omega\,\bar{\Omega})^2} \left[ (1+\Omega\, 
\bar{\Omega})\,\frac{dt}{a} - (1-\Omega\,\bar{\Omega})\,d\chi + 
\frac{2\,\Sigma}{1+\xi\bar{\xi}}\left(\bar{\Omega}\,d\xi + \Omega\, 
d\bar{\xi}\,\right) \right]^{2} \, .  \label{gksm} 
\end{eqnarray} 
It is interesting to point out that this family of stationary  
solutions of Einstein's equations has an arbitrary function, namely  
$\Omega$.  Moreover, as far as we know, these metrics were  
previously unknown. 
 
With regard to the complex function $\Omega$, if the function  
$G(z_{1},z_{2})$ does not depend on $z_{2}$ we have 
\[ \Omega = \mbox{e}^{\sqrt{-\varepsilon\,\chi}}\, 
f(\xi,\bar{\xi}) \, , \] 
where $f$ is a complex function that we can find by solving the  
system of differential equations (\ref{oec1}--\ref{oec2}). This  
form of $\Omega$ implies that 
\[ \hh = \left\{ \begin{array}{ll} 0 & \mbox{if}\hspace{3mm} 
\varepsilon=0\, , 1\, ,\\ \frac{\sqrt{2}\,m}{a} & \mbox{if} 
\hspace{3mm} \varepsilon=-1 \, , \end{array} \right.  \] 
and this means that in order to avoid $\hh = 0$ for the cases 
$\varepsilon = 0,1$ (which would mean that the GKS transformation  
is trivial), we must impose that 
\[ \frac{\partial\,G}{\partial z_{2}} \neq 0 \, , \] 
and therefore, by the implicit function theorem, we can write 
(\ref{fung}) as follows 
\[ \Upsilon\left( \frac{\mbox{e}^{-\sqrt{-\varepsilon\,\chi}}\, 
\Omega - \xi}{1 + \bar{\xi}\,\mbox{e}^{-\sqrt{-\varepsilon\,\chi}} 
\,\Omega}\,\right)\hspace{1mm}+\hspace{1mm}\frac{(1-\xi\bar{\xi}) 
\,\Sigma'\,\Omega - \xi + \bar{\xi}\,\Omega^{2}}{(1+\xi\bar{\xi}) 
\,\Sigma\,\Omega}  = 0 \, , \] 
where $\Upsilon(z)$ is any analytic complex function of one complex  
variable $z$.  Moreover, as we can see directly from the last  
equation, or from equations (\ref{oec1}--\ref{oec3}) and the  
expression for the line--element (\ref{gksm}), the transformation 
\[ \Omega \longrightarrow \Omega' = - \frac{1}{\bar{\Omega}} , 
\hspace{1cm} \chi \longrightarrow \chi' = - \chi \, , \] 
is a {\em gauge} transformation since it induces the change  
$\vec{\tilde{\ell}}\hspace{1mm}\longrightarrow\hspace{1mm} 
\vec{\tilde{\ell}}{}^{'} = - \vec{\tilde{\ell}}$, and  
therefore both solutions of equations (\ref{oec1}--\ref{oec3})  
$\Omega$ and $\Omega'$ determine the same GKS metric.  Apart from this  
gauge, we have not been able to find any other.  On the other hand,  
we can think, as Debney \cite{DEBN} did, that part of the complex  
functions $\Omega$ are gauges of the seed metric, or in other words,  
that the GKS metrics we obtain are the seed metrics.  Looking at the  
expressions (\ref{psi0}--\ref{psi4}) for the Weyl tensor, this can  
only happen when $\rho+\bar{\rho}=0 \Rightarrow 
\bar{\delta}\rho=0$, which corresponds to case A). Therefore, 
the GKS metrics (\ref{gksm}) are always different from the seed  
metrics.  
 
As we said before, the family of solutions (\ref{gksm}) is too wide  
in the sense that it depends on the arbitrary function $\Omega$.   
Analyzing the form of the metric (see also \cite{SOPU}) the solutions  
contained in each case are: 
 
\vspace{2mm} 
 
$\bullet\,$ Case $\varepsilon=0$: If we do not consider the  
cosmological constant ($C=0$), the metrics (\ref{gksm}) are all the  
classical vacuum Kerr--Schild metrics \cite{DEKS} with the exception  
of the plane--fronted gravitational waves \cite{TRAU,DEBN,URBA}.  
This exception is due to the fact that this class of solutions  
corresponds to the case $\rho = 0$ (see \cite{KSHM}).  The  
Schwarzschild spacetime is obtained when $Y=\Omega=0$, and the Kerr  
spacetime when we choose $\Upsilon$ in the following way 
\begin{equation} 
\Upsilon = - \frac{1}{\sqrt{2}\,\mbox{i}\,c} \, , \label{keva} 
\end{equation} 
where $c$ is an arbitrary real constant. In this case, if we carry  
out the suitable changes (see \cite{SOPU}), we can identify the mass  
and the angular momentum as measured from infinity with  
$m/(4\,\sqrt{2}\,a)$ and  $(m\,c)/(4\,\sqrt{2}\,a)$ respectively.  
 
\vspace{2mm} 
 
$\bullet\,$ Case $\varepsilon=1$: All the  
spacetimes within this case were unknown with the exception of the  
case given by (\ref{keva}), corresponding to a metric due to  
Vaidya \cite{VAID} (it is also a particular case of the Wahlquist  
metric \cite{WAHL,HEHE}) which has been interpreted  
as the Kerr metric in the cosmological background of the Einstein  
Universe. In this sense, the metrics of this class can be  
interpreted as the vacuum Kerr--Schild configurations in the  
background of the Einstein Universe.  
 
\vspace{2mm} 
 
$\bullet\,$ Case $\varepsilon=-1$: These spacetimes were also unknown  
and they correspond to the Kerr--Schild configurations in the  
background of the seed spacetimes (\ref{seed}) with $\varepsilon=-1$.   
From the physical point of view these spacetimes have less interest  
since  they do not satisfy the usual energy conditions \cite{HAEL}. 
 
\section{Solution of the Killing equations and study of the  
symmetries} \label{sec6} 
\mbox{}\indent  
The subject of this section is the study of the symmetries of the  
GKS spacetimes (\ref{gksm}).  A remarkable point in this study is  
the fact that it is possible to integrate the Killing equations without the  
explicit use of coordinates, as it happened   
with the equations for $\hh$.  To carry out this integration, we  
consider the Killing equations for these metrics 
\begin{equation} 
\tilde{\nabla}_{a}\,\zeta_{b}\hspace{1mm}+\hspace{1mm} 
\tilde{\nabla}_{b}\,\zeta_{a}\hspace{1mm}=\hspace{1mm} 0 \, ,  
\label{kill} 
\end{equation} 
where $\vec{\zeta}$ is a hypothetical Killing vector field for the 
metrics (\ref{gksm}), and  we project them onto the null basis  
$\{\ell,\kk,\mm,\bar{\mm}\}$ associated with the seed spacetimes,  
whiat allows us to use its spin coefficients and their properties  
(\ref{coes}). 
 
After making the integration (see \cite{SOPU} for details), we  
find that in the general case, that is to say, without additional  
assumptions on the complex function $\Omega$, there is only one  
independent Killing vector field. Obviously, this Killing vector  
field is the timelike one that we already know 
\begin{equation} 
\vec{\zeta} = \frac{1}{\sqrt{2}}\,\left[\,(1-\hh)\, 
\vec{\tilde{\ell}} + \vec{\tilde{\kk}}\,\right] = \frac{1} 
{\sqrt{2}}\,\left(\vec{\ell} + \vec{\kk}\,\right) = \vec{u}  
\propto \vec{\tilde{u}} \, . \label{zeta}  
\end{equation} 
Thus, the solution (\ref{gksm}) is stationary in general with  
\underline{no} further symmetries.  The norm of this Killing vector  
field is $\hh-1$, which for the cases $\varepsilon\pm 1$ is always  
negative [otherwise we cannot define $\vec{\tilde{u}}$  
(\ref{vecu})], but in the case $\varepsilon=0$ there are not  
restrictions and then, the sign of the norm divides the spacetime into  
three regions: A) $1-\hh>0$, where the Killing is timelike. B)  
$1-\hh<0$, where the Killing is spacelike. C) $1-\hh=0$, where the  
Killing is null. In fact, that is the stationary limit hypersurface 
(this can be checked easily for the Schwarzschild and Kerr  
spacetimes). This division according to the norm of the Killing vector 
field (\ref{zeta}) is an interesting fact because these three  
regions are usually seen as different spacetimes (see \cite{KSHM}). 
 
Apart from the general case, there are specific cases with  
additional symmetries.  These cases appear when the quantity  
$\Gamma+\bar{\Gamma}$ vanishes, where $\Gamma$ is a complex function  
defined by 
\begin{equation} 
\Gamma \equiv \frac{1}{(\bar{\delta}\rho)^{2}}\left\{ (\rho^{2}+ 
\frac{\varepsilon}{2\,a^2})(\bar{\delta}\bar{\delta}\rho + 2\,\alpha 
\bar{\delta}\rho) - 3\,\rho(\bar{\delta}\rho)^{2} \right\} \, .  
\label{gmgb} 
\end{equation} 
 
Now, two different cases appear depending on whether the complex  
function $\Gamma$ vanishes or not. 
 
\vspace{3mm} 
 
\underline{Case (i)}: $\Gamma \neq 0$.  After integrating the  
Killing equations (\ref{kill}) we find that there are two  
independent Killing vector fields, which we can take as  
$\zeta^{a}$, given in (\ref{zeta}), and $\zeta_{1}^{a}$, given by 
\begin{equation} 
\vec{\zeta}_{1} =  \frac{1}{\sqrt{\mid\Gamma\mid}}\left\{ (1+\hh)\, 
\vec{\tilde{\ell}}-\vec{\tilde{\kk}}-\frac{2\,(\bar{\rho}^{2}+  
\frac{\varepsilon}{2\,a^2})}{\delta\bar{\rho}}\,\vec{\tilde{\mm}}- 
\frac{2\,(\rho^{2}+\frac{\varepsilon}{2\,a^2})} 
{\bar{\delta}\rho}\,\vec{\tilde{\bar{\mm}}}\right\} \, .  
\label{kiib} 
\end{equation} 
It is clear that we can write the most general Killing vector  
field as follows  
\begin{equation} 
C_{1}\,\vec{\zeta} + C_{2}\,\vec{\zeta}_{1} \, , \label{geex} 
\end{equation} 
where $C_{1}$ and $C_{2}$ are two arbitrary constants. It can be  
shown that this general  Killing vector field commutes with the  
timelike one $\vec{\zeta}$ given in (\ref{zeta}).  Furthermore, we  
can study the possible existence of an axial Killing vector field.  
To that end we must check the regularity condition  
\begin{equation} 
\frac{\tilde{\nabla}^{a}\,\zeta^{2}\,\tilde{\nabla}_{a}\, 
\zeta^{2}}{4\,\zeta^{2}} \hspace{3mm} \stackrel{\mbox{{\tiny axis}}} 
{\longrightarrow}\hspace{3mm} 1 \, , \hspace{10mm} \zeta^{2}  
\equiv \zeta^{a}\,\zeta_{a} \, . \label{axis} 
\end{equation} 
For the general Killing vector field (\ref{geex}) this condition  
is not satisfied for any value of $C_{1}$ and $C_{2}$ and therefore, 
there is no axial symmetry in this case. 
 
\vspace{5mm} 
 
\underline{Case (ii)}:  $\Gamma = 0$.  In this case, equation 
(\ref{petr}) implies that the GKS spacetimes (\ref{gksm}) are 
Petrov type D. When we integrate the Killing equations (\ref{kill}),  
we find that there are also two independent Killing vector fields,  
which can be taken as (\ref{zeta}) and  
\begin{eqnarray} 
\vec{\zeta}_{2} = \frac{1}{(\rho^{2} + \frac{\varepsilon}{2\, 
a^2})^{\frac{3}{2}}(\bar{\rho}^{2} + \frac{\varepsilon}{2\,a^2})^{ 
\frac{3}{2}}}\left\{\bar{\delta}\rho\,\delta\bar{\rho}\,\left[\, 
(1+\hh)\,\vec{\tilde{\ell}}-\vec{\tilde{\kk}} \, \right] \right. - 
\hspace{3cm} \mbox{}  \nonumber  \\ 
\mbox{} \hspace{5cm} \left. 2\,(\bar{\rho}^{2}+\frac{\varepsilon} 
{2\,a^2})\, \bar{\delta} \rho\,\vec{\tilde{\mm}} - 2\,(\rho^{2}+ 
\frac{\varepsilon}{2\,a^2})\, \delta\bar{\rho}\, 
\vec{\tilde{\bar{\mm}}} \right\} \, . \label{kild} 
\end{eqnarray} 
In the same way that in the case (i), we can check that the general  
expression for a Killing vector field $C_{1}\,\vec{\zeta} + C_{2}\, 
\vec{\zeta}_{2}$, where $C_{1}$ and $C_{2}$ are two arbitrary  
constants, commutes with the timelike Killing vector field  
$\vec{\zeta}$.  In addition, and with regard to the possibility of  
axial symmetry, it can be shown that we can only have a spacelike  
Killing vector field vanishing in a certain region of the spacetime  
when $C_{1}=0$, and then, this region is defined by $\mid \!  
\bar{\delta}\rho \! \mid = 0$.  Moreover, it can be checked that if  
the following relation holds 
\begin{equation} 
\left. (\rho-\bar{\rho})^{-2}\,(\rho^{2}+\frac{\varepsilon} 
{2\,a^2})\,(\bar{\rho}^{2}+\frac{\varepsilon}{2\,a^2})  
\right|_{\mid \bar{\delta}\rho \mid \rightarrow 0} \hspace{3mm}  
\longrightarrow \hspace{3mm}\mbox{const.} \, , \label{poax} 
\end{equation} 
and we choose the constant $C_{2}$ as follows 
\[ \left. C_{2} =  \sqrt{\frac{(\rho^{2}+\frac{\varepsilon} 
{2\,a^2})(\bar{\rho}^{2}+\frac{\varepsilon}{2\,a^2})}{-8(\rho- 
\bar{\rho})^{2}}}\,\right|_{\mid \bar{\delta}\rho \mid  
\rightarrow 0} \, , \] 
the axis regularity condition (\ref{axis}) is satisfied for the  
Killing vector field $C_{2}\,\vec{\zeta}_{2}$, and therefore, it  
is an axial Killing vector field. In that case, the GKS spacetimes  
would be stationary axially symmetric spacetimes in {\em rigid  
rotation} because $\vec{\tilde{u}}$ is proportional to the timelike  
Killing vector field (\ref{zeta}). For the Kerr and the  
Kerr--Vaidya metrics, which are given by (\ref{keva}) and  
$\varepsilon = 0,1$ respectively, it can be checked that the  
condition (\ref{poax}) is fulfilled.  In general, following  
\cite{SEN2}, we can conclude that all the spacetimes in this case  
with axial symmetry are subcases of the Wahlquist family  
\cite{WAHL}. 
 
\vspace{5mm} 
 
\noindent{\Large {\bf Acknowlegments}} 
 
\vspace{5mm} 
 
I would like to express my gratitude to Dr. J.M.M. Senovilla for  
helpful discussions, many valuable suggestions and for his  
collaboration in the preparation of this manuscript. Some calculations  
have been performed with the help of the computer programme REDUCE.

\appendix 
 
\section{Conformally--flat form of the FLRW spacetimes} \label{appa} 
\mbox{}\indent 
In this appendix we are going to give a coordinate change to pass 
from the following well--known form for the line element of the 
FLRW spacetimes 
\begin{equation} 
ds^2 = -dt^2+a^2(t) \left\{ d\chi^2 + \Sigma^2(\varepsilon,\chi)\,( 
d\theta^2 + \sin^2\theta d\varphi^2  ) \right\} \, , \label{lerw} 
\end{equation} 
where $\varepsilon^3=\varepsilon$ and $\Sigma(\varepsilon,\chi)$  
satisfies the differential equation 
\begin{eqnarray*} 
\Sigma'^2 + \varepsilon\,\Sigma^2 = 1 \hspace{3mm} \Longleftrightarrow 
\hspace{3mm} \Sigma(\varepsilon,\chi) = \left\{ \begin{array}{ll} 
 \sin\chi    & \mbox{if $\varepsilon = 1$\, ,}  \\ 
 \chi       & \mbox{if $\varepsilon = 0$\, ,}  \\ 
 \sinh\chi   & \mbox{if $\varepsilon = -1$\, ,} \end{array} \right. 
\end{eqnarray*} 
to the following explicitly conformally--flat form 
\[ds^2 = \Phi^2(T,R)\left\{-dT^2+dR^2+R^2 (d\theta^2+ \sin^2\theta  
d\varphi^2) \right\} \, . \] 
This transformation can be achieved by using the following coordinate 
change 
\[ T = \frac{2\,\Sigma(\varepsilon,\psi)}{\Sigma'(\varepsilon,\psi) 
+\Sigma'(\varepsilon,\chi)} \, , \hspace{5mm} R = \frac{2\,\Sigma( 
\varepsilon,\chi)}{\Sigma'(\varepsilon,\psi)+\Sigma'(\varepsilon,\chi)}  
\, , \] 
where $\psi$ is the so--called {\em parametric time}, defined by 
\[ \dot{\psi} = \frac{d\psi}{dt} \equiv  \frac{1}{a} \, . \] 
After making this coordinate change, the conformal factor $\Phi$ becomes 
 
\[ \Phi^2 = \frac{a^2}{\left[1+\varepsilon\left(\frac{T-R}{2}\right)^{2} 
\right]\left[1+\varepsilon\left(\frac{T+R}{2}\right)^{2}\right]} . \] 
In these coordinates the fact that the scale factor $a$ depends only 
on $t$ reads as follows 
\[ a \hspace{3mm} = \hspace{3mm} a\left(\frac{T}{1-\frac{\varepsilon} 
{4}(T^2 - R^2)} \right) \, . \] 
 
Another interesting form for the line element (\ref{lerw}) is obtained 
by passing to the following null spherical coordinates, 
\[ U = \frac{T-R}{2}\, , \hspace{1cm} V = \frac{T+R}{2}\, , \hspace{1cm} 
\xi = e^{i\varphi}\tan\left(\frac{\theta}{2}\right) \, , \] 
then, we obtain 
\[ ds^2 = \Phi^2(U,V)\left\{-4 dU dV + 4\left(\frac{V-U}{1+\xi\bar{\xi} 
}\right)^{2} d\xi d\bar{\xi} \right\}\, , \hspace{3mm} \Phi^2(U,V)= 
\frac{a^{2}}{(1+\varepsilon\,U^2)(1+\varepsilon\,V^2)} \, , \] 
where 
\[ a \hspace{3mm} = \hspace{3mm} a\left(\frac{V+U}{1-\varepsilon\,U\,V} 
\right) \, .\] 
 
\section{Kinematical quantities and the Newman--\-Penrose \newline formalism} 
\label{appb}\mbox{}\indent 
In this appendix we give some useful formulae due to Wainwright 
\cite{WAIN}, for the kinematical quantities of a unit timelike 
vector field.  Let $\vec{u}$ be any unit timelike vector field and 
let $\{\ell,\kk,\mm,\bar{\mm}\}$ be a null basis such that we can 
write down $\vec{u}$ in the following way 
\[ \vec{u} \hspace{2mm} = \hspace{2mm} \frac{1}{\sqrt{2}} 
(\vec{\ell}+\vec{\kk}) \, . \] 
Then, we can write the kinematical quantities associated with  
$\vec{u}$ using only the null basis $\{\ell,\kk,\mm,\bar{\mm}\}$  
and the spin coefficients associated with them as follows 
 
\vspace{3mm} 
 
\noindent {\em Expansion} 
\[ \theta = \frac{1}{\sqrt{2}}\left(\epsilon+\bar{\epsilon}-\gamma- 
\bar{\gamma}+\mu+\bar{\mu}-\rho-\bar{\rho} \right) \, , \] 
 
\vspace{5mm} 
 
\noindent {\em Acceleration} 
\[ \vec{\mbox{\cal{a}}} = \frac{1}{\sqrt{2}}\left(\epsilon+ 
\bar{\epsilon}+\gamma+\bar{\gamma}\right)\vec{v} + \frac{1}{2}\left 
\{\left(\pi-\bar{\tau}+\nu-\bar{\kappa}\right)\vec{\mm}+\mbox{c.c.}  
\right\} \, , \] 
 
\vspace{5mm} 
 
\noindent {\em Shear} 
\begin{eqnarray*} 
\sigma_{ab} & = & \frac{1}{3\sqrt{2}}\left\{2(\epsilon+ 
\bar{\epsilon})+\rho+\bar{\rho}-2(\gamma+\bar{\gamma})-\mu-\bar{\mu} 
\right\}(v_{a}v_{b}-\mm_{(a}\bar{\mm}_{b)}) +  \\ 
& & \frac{1}{2}\left\{\left(2(\alpha+\bar{\beta})+\pi+\bar{\tau}-\nu- 
\bar{\kappa}\right)v_{(a}\mm_{b)} + \mbox{c.c.} \right\} + 
\frac{1}{\sqrt{2}}\left\{\left(\lambda-\bar{\sigma}\right)\mm_{a} 
\mm_{b}+\mbox{c.c.} \right\} \, , 
\end{eqnarray*} 
 
\vspace{5mm} 
 
\noindent {\em Rotation} 
\[ \omega_{ab} = \frac{1}{2}\left\{\left(2(\alpha+\bar{\beta})+\nu+ 
\bar{\kappa}-\pi-\bar{\tau}\right)v_{[a}\mm_{b]} + \mbox{c.c.}  
\right\}+\frac{1}{\sqrt{2}}\left(\rho-\bar{\rho}+\mu-\bar{\mu} 
\right)\mm_{[a}\bar{\mm}_{b]} \, , \] 
where 
\[ \vec{v} \hspace{2mm} \equiv \hspace{2mm} \frac{1}{\sqrt{2}} 
(\vec{\ell}-\vec{\kk}) \hspace{5mm} \Longrightarrow \hspace{5mm} 
v^{a}v_{a} = 1 \, , \hspace{5mm} v^{a}u_{a} = 0 \, . \] 
 
\pagebreak

\end{document}